# Theoretical Investigation of the Magnetoelectric Properties of $Bi_2NiTiO_6$

Lokanath Patra[1, 2] and P. Ravindran[1, 2, 3, 4, a)]

[1]*Department of Physics, Central University of Tamil Nadu, Thiruvarur-610005.*
[2]*Simulation Center for Atomic and Nanoscale Materials, Central University of Tamil Nadu, Thiruvarur-610005.*
[3]*Department of Materials Science, Central University of Tamil Nadu, Thiruvarur-610005.*
[4] *Department of Chemistry, Center for Materials Science and Nanotechnology, University of Oslo, P.O. Box 1033 Blindern, N 1035 Oslo, Norway*

[a)]Corresponding author: raviphy@cutn.ac.in

**Abstract.** We report the first principle investigations on the structural, electronic, magnetic and ferroelectric properties of a Pb free double perovskite multiferroic $Bi_2NiTiO_6$ using density functional theory within the general gradient approximation (GGA) and GGA+$U$ method. Our results show that $Bi_2NiTiO_6$ will be an insulator with $G$-type magnetic ordering in its ground state with $Ni^{2+}$ in a high spin state and a spin moment of $1.74\mu_B$. The paraelectric phase stabilizes in nonmagnetic state with $Ni^{2+}$ in low spin configuration showing that spin state transition plays an important role in strong magnetoelectric coupling in $Bi_2NiTiO_6$. The bonding characteristics of the constituents are analyzed with the help of partial density of states and Born effective charges. The presence of Ti ions at Ni sites suppresses the disproportionation observed in case of $BiNiO_3$ and results in a noncentrosymmetric crystal structure. The coexistence of Bi 6$s$ lone pair and $Ti^{4+}$ $d^0$ ions which brings covalency produces a polarization of 32 $\mu Ccm^{-2}$.

## INTRODUCTION

Bi- and Pb- based multiferroic compounds have received much attention in recent years due to their high value of spontaneous electrical polarization[1–3]. It is believed that the spontaneous polarization in these compounds mainly results from the n$s^2$ lone pair which gives off-center displacement of ions and as a result noncentrosymmetry to the crystal. But there has been growing interest in Bi-based multiferroics due to the toxic nature of Pb. $BiNiO_3$ has been found to be an insulating antiferromagnet which is heavily distorted triclinic symmetry[4]. But due to the centrosymmetry nature of the compound, it lacks spontaneous electrical polarization. This is because the oxidation state was found to be $Bi_{1/2}^{3+}Bi_{1/2}^{5+}Ni^{2+}O_3^{2-}$ rather than $Bi^{3+}Ni^{3+}O_3^{2-}$. So, the system may be substituted with suitable cation which would break the centrosymmetric nature as well as suppress the disproportionation of Bi ions. Another way of getting enhanced polarization is the presence of $d^0$ ions in the system. For example, in $BaTiO_3$ due to the covalent bonding between $Ti^{4+}$ ($d^0$ ion) and $O^{2-}$, the Ti atom gets displaced position to produce a ferroelectric ground state structure[5]. Therefore, a possible strategy to get a multiferroic compound from $BiNiO_3$ is a chemical modification of the compound with a $d^0$ ion at the *B*-site and that motivated the current study. Hence, we have performed the first principle investigation on the multiferroic properties of $Bi_2NiTiO_6$, which has been studied as ferroelectric materials recently[6]. However, there has not yet been theoretical calculations about the structure, electronic, ferroelectric and magnetic properties of $Bi_2NiTiO_6$, though there are a few experimental reports available[6–8].

## COMPUTATIONAL DETAILS

The calculations are performed by VASP[9] code within the generalized gradient approximation (GGA)[10]. In the present study, we have used GGA with the Perdew-Burke-Ernzerhof (PBE)[11] as exchange and correlation functional. To properly describe the strong correlation of the Ni 3$d$ electrons, we have used GGA+$U$ method. Here $U_{eff} = U - J$ was used instead of *U*. The value of $U_{eff}$ = 6 eV was used throughout the calculations because it was found to be sufficient for our previous work on nickelates[12]. As ferroelectric properties are very sensitive to structural parameters, we have used a very high cut-off energy of 800 eV to reproduce the structural parameters correctly. The calculations were carried out with 7x5x7 Monkhorst-pack[13] **k**-point mesh. Convergence is assumed to be attained when the energy difference between two successive iterations is less than $10^{-6}$ eV per cell and the forces acting on the atoms are less than 1meVÅ$^{-1}$. To find the ground state magnetic ordering we have performed total energy calculations for paramagnetic (PM), ferromagnetic (FM), *A*-, *C*- and *G*- type antiferromagnetic (*A*-AFM, *C*-AFM, *G*-AFM) configurations[12]. The Born-effective charges (BEC) were calculated using the so-called "Berry phase finite difference approach"[14].

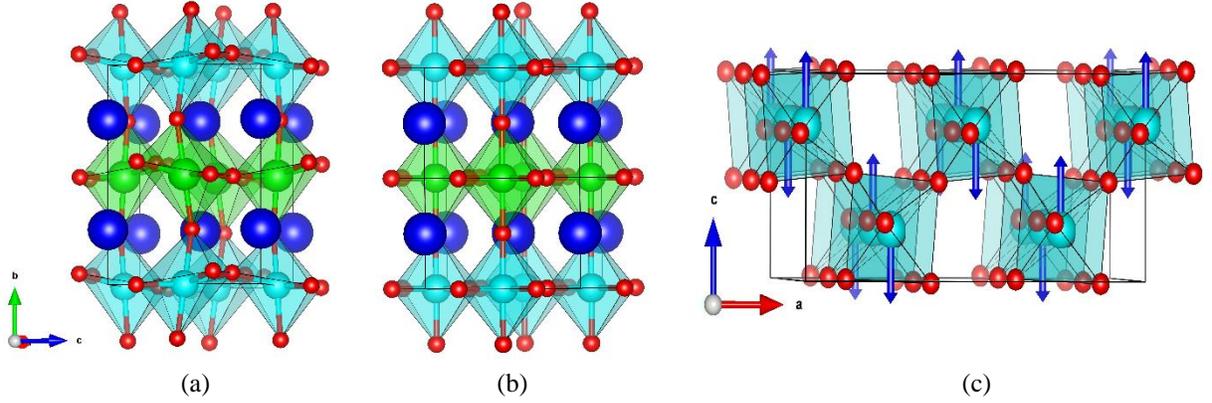

(a) (b) (c)

**FIGURE 1.** The optimized crystal structure of $Bi_2NiTiO_6$ in (a) noncentrosymmetric *Pn21a* space group, (b) centrosymmetric *Pnma* space group, (c) The 2x2x1 supercell of the ferroelectric ground state with *G*- type magnetic ordering (only Ni-O octahedras are shown). The blue, cyan, green and red spheres represent Bi, Ni, Ti, and O, respectively.

## RESULTS AND DISCUSSIONS

The introduction of $Ti^{4+}$ ions in $BiNiO_3$ lattice breaks the centrosymmetric nature and thus produces a ferroelectric ground state and hence $Bi_2NiTiO_6$ crystallizes in an orthorhombic structure with space group *Pn21a*. The atom positions are optimized for the experimental crystal structure[6]. Moreover, in order to find the optimized ground state global minima, we have adopted force as well as stress minimization method as implemented in VASP. The optimized volume was found to be 248.44 Å$^3$ which is overestimated by only 1.1% compared with the experimental value (245.78 Å$^3$)[6]. The optimized (experimental) lattice parameters are $a$ = 5.74 (5.61) Å, $b$ = 7.88 (7.85) Å, and $c$ = 5.49 (5.58) Å and are in good agreement with experimental values. The Bulk modulus and its pressure derivative were found to be 120.87 (115) GPa and 7.14, respectively. The optimized crystal structure is given in Fig. 1(a). The crystal structure consists of Ni/Ti-O distorted corner shared octahedra. It can be seen that Bi, Ni/Ti ions have shifted away from their positions in the centrosymmetric space group *Pnm*a given in Fig. 1(b).

Our total energy calculations reveal that $Bi_2NiTiO_6$ stabilizes with *G*-AFM ordering as shown in Fig. 1(c). The ground state *G*-AFM configuration is lower in energy than FM, *A*-AFM and *C*-AFM by 158 meV, 99 meV and 57 meV, respectively. The magnetic moment at Ni site for the ground state *G*-AFM configuration was found to be 1.74 $\mu_B$. In an ideal octahedral cubic crystal field, the *d*-level splits into triply degenerate $t_{2g}$ ($d_{xy}$, $d_{xz}$, $d_{yz}$) and doubly degenerate $e_g$ ($d_{x^2-y^2}$, $d_{z^2}$) levels. In pure ionic case, $Ni^{2+}$ with 8 *d*-electrons will fill these energy levels with a low spin (LS) state ($t_{2g}^6$, $d_{x^2-y^2}^2$, $d_{z^2}^0$) or high spin (HS) state ($t_{2g}^6$, $d_{x^2-y^2}^1$, $d_{z^2}^1$) with spin moments 0 $\mu_B$ and 2 $\mu_B$, respectively. As it can be seen from our orbital projected DOS for ground state *G*-AFM configuration given in Fig. 2(a), the $t_{2g}$ orbitals are fully occupied. But the $e_g$ orbitals are occupied only in the majority spin channel which shows that the $Ni^{2+}$ ions are in HS spin state. The orbital projected DOS distribution shows nearly empty orbitals for $t_{2g}$ states of Ti *d* and finite DOS can be seen at $e_g$ states due to covalency effect. From this illustration, it is confirmed that Ti atom tends to form an oxidation state of 4+. Moreover, the centrosymmetric structure is found to be nonmagnetic with $Ni^{2+}$ in LS state which indicates that spin state transition plays vital role on strong magnetoelectric coupling in $Bi_2NiTiO_6$. Hence, ferroelectric to paraelectric phase transition can be achieved with applied external magnetic field that changes the spin state of $Ni^{2+}$ i.e. HS to LS and with the application of an external electric field one can even go from magnetic to non-magnetic state in $Bi_2NiTiO_6$. Similar results were found for our previous work on $BiCoO_3$[1].

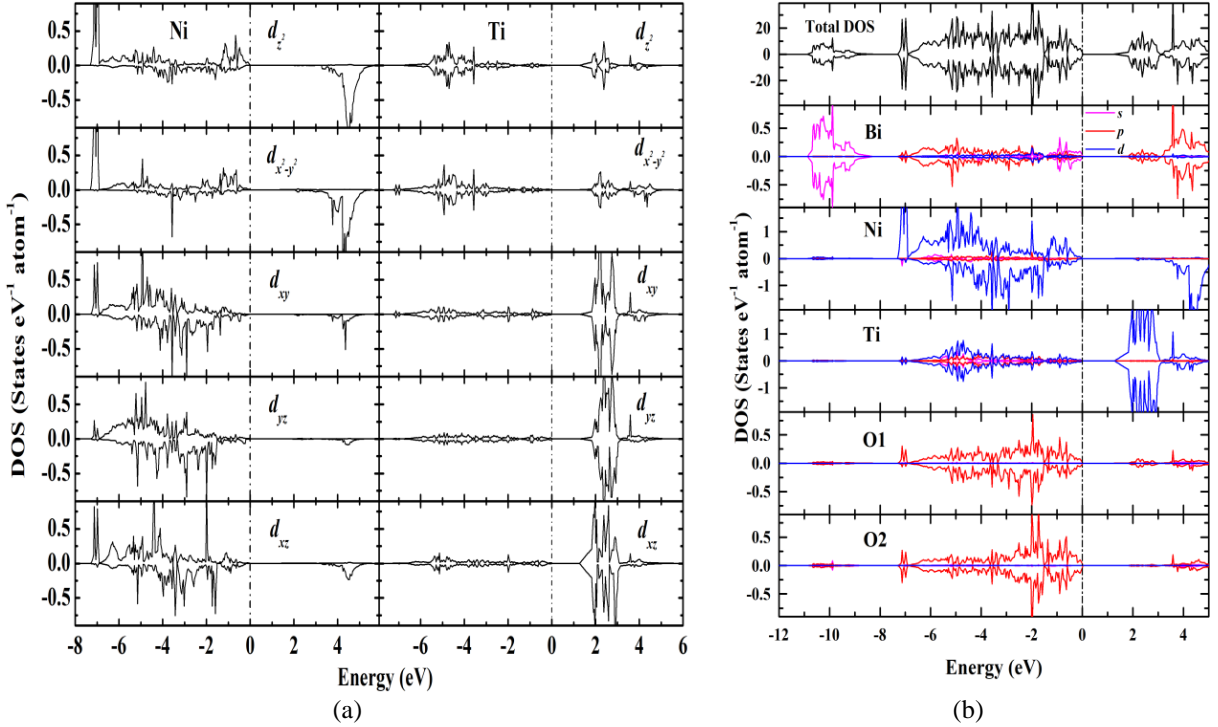

**FIGURE 2.** (a) Orbital projected DOS, (b) Total and partial DOS for $Bi_2NiTiO_6$ in *G*-AFM configuration using GGA+*U* method with $U_{eff} = 6$

Figure 2(b) shows the total and partial DOS for $Bi_2NiTiO_6$ for the ground state *G*-AFM configuration. The top of the valence band and the bottom of the conduction band are separated from each other by an energy gap of 1.4 eV resulting in an insulating behavior. The partial DOS given in Fig. 2(b) shows that the DOS for Bi 6*p*, Ni 3*d*, Ti 3*d* and O 2*p* orbitals are degenerate in the energy range from -7.5 to Fermi level ($E_F$) which ensures the covalent bonding between these atoms. Due to the covalent interaction between Ti 3*d* and O 2*p* states, the charges redistribute between these states resulting in the occupation of Ti *d* states as evident in Fig.2(b). Moreover, due to these strong covalent hybridization, the calculated total DOS curves given in Fig. 2(b) show broad features. As covalent bonding is present between Ni 3*d* and neighboring O 2*p* sites, one can expect induced magnetic moment at the O sites. Our GGA+*U* calculation shows that the magnetic moment present at the oxygen sites is ~ 0.02 $\mu_B$. Further, this covalent interaction reduces the magnetic moment at the Ni site from the moment obtained from pure ionic picture. Hybridization between Ni 3*d* and the surrounding oxygen ligands is well known to lead to superexchange interactions in magnetic perovskites. From the Bi partial DOS, a sharp peak can be seen around -10 eV which originates from the Bi 6*s* lone pair electrons.

The Born effective charges calculated with Berry phase method are listed in Table 1. In $Bi_2NiTiO_6$, the formal valencies of Bi, Ni, Ti, and O are 3+, 2+, 4+ and 2-, respectively. But the charges on Bi, Ni, Ti, and O are larger than the nominal ionic value which reveals the presence of a large dynamic contribution superimposed to the static charge. In high symmetry oxides, the Born effective charge tensor ($Z^*$) for O is isotropic and close to -2. But, as $Bi_2NiTiO_6$ crystallizes in low symmetry orthorhombic *Pn21a* space group, owing to the site symmetry the diagonal components of $Z^*$ are non-integer for all the ions listed in Table 1. The finite off-diagonal components at the oxygen sites confirm the covalent bonding between O 2*p* and Ni 3*d* orbitals. The BEC can also be used to quantify polarization in $Bi_2NiTiO_6$.

It is well known that the paraelectric phase produces zero net polarization due to the absence of off-center displacement. So, the spontaneous electric polarization can be calculated by keeping the optimized lattice parameters and atom positions of both the paraelectric and ferroelectric phase and calculate the displacements of ions with respect to the paraelectric phase. As we know the BEC from our Berry phase calculation, the multiplication of ionic displacements with BEC will give the polarization contribution from each ion in the system. The summation of polarization for all the ions will give the net spontaneous polarization present in the ferroelectric phase. In $Bi_2NiTiO_6$, the Bi, Ni, and Ti ions all shift away from the positions of (0, 0.25, 0), (0, 0, 0.5), and (0, 0, 0.5) in the nearest centrosymmetric *Pnma* space group to form distorted octahedras as shown in Fig. 1. The calculated spontaneous polarization for $Bi_2NiTiO_6$ was found to be ~ 32 $\mu Ccm^{-2}$ which is comparable to the polarization of conventional ferroelectric $BaTiO_3$. Our partial polarization analysis shows that the polarization not only coming from the contribution from $Bi^{3+}$ lone pairs but also due to the displacement of ions when going from paraelectric to ferroelectric phase due to strong Bi-O and Ti-O covalent hybridization. Similar calculations using the nominal charges produced a polarization of ~ 19 $\mu Ccm^{-2}$, almost 40% less than the value obtained using the BECs and this reduction in polarization is related to the covalency effect.

**TABLE 1.** Calculated Born effective charge tensor for antiferromagnetic $Bi_2NiTiO_6$ in the ferroelectric $Pn2_1a$ structure

| Ion | $Z_{ij}^*$ | | | | | | | | |
|---|---|---|---|---|---|---|---|---|---|
| | xx | yy | zz | xy | yz | zx | xz | zy | yx |
| Bi1 | 4.976 | 5.358 | 5.523 | -0.054 | 0.098 | -0.056 | 0.071 | 0.339 | -0.485 |
| Bi2 | 4.976 | 5.358 | 5.537 | -0.053 | 0.098 | -0.057 | 0.074 | 0.338 | -0.485 |
| Ni | 1.886 | 1.857 | 2.106 | -0.241 | -0.044 | -0.014 | 0.037 | 0.195 | 0.064 |
| Ti | 7.072 | 7.143 | 6.211 | 0.198 | -0.602 | 0.224 | -0.395 | 0.336 | 0.615 |
| O1 | -2.499 | -3.250 | -3.742 | 0.641 | 0.763 | 0.303 | 0.287 | 0.711 | 0.472 |
| O2 | -2.497 | -3.250 | -3.749 | 0.635 | 0.763 | 0.300 | 0.280 | 0.707 | 0.472 |
| O3 | -2.656 | -2.543 | -3.476 | -0.627 | -0.577 | 0.013 | 0.097 | -0.453 | -0.729 |

## CONCLUSION

We have shown here that magnetoelectric multiferroics can be obtained from paraelectric phases of perovskites by suitable cationic substitution. From the spin polarized total energy calculations we found that the ground state of $Bi_2NiTiO_6$ stabilizes in $G$-type magnetic ordering with a bandgap of 1.4 eV. Our orbital projected DOS analysis finds that $Ni^{2+}$ is in a high spin state with a spin moment of 1.74 $\mu_B$. The paraelectric phase stabilizes with nonmagnetic phase where $Ni^{2+}$ is in LS state and in contrast the ferroelectric phase is found to be magnetic with $Ni^{2+}$ is in HS state. We have once again shown that magnetic instability can bring giant magnetoelectric coupling and in extreme cases one can go from magnetic to nonmagnetic vice versa by applying electric field as shown here. Detailed analysis on partial DOS and BEC shows strong covalent interaction between transition metals and oxygen. the ferroelectric phase stabilizes by the lone pair electron at the $Bi^{3+}$ site and covalency from $Ti^{4+}$ ion with a polarization of 32 $\mu Ccm^{-2}$.

## ACKNOWLEDGMENTS

The authors are grateful to the Research Council of Norway for providing computing time at the Norwegian supercomputer facilities. This research was supported by the Indo-Norwegian Cooperative Program (INCP) via Grant No. F. No. 58-12/2014(IC)